\documentclass[reprint, prl,twocolumn,showpacs,superscriptaddress,floatfix]{revtex4-1}

\usepackage{epsfig}
\usepackage{graphics} 
\usepackage{makeidx}
\usepackage{colordvi}
\usepackage{amsmath}
\usepackage[percent]{overpic}
\usepackage{subfigure}
\usepackage{mdwlist}

\RequirePackage{xspace}
\usepackage{relsize}

\def\pep2{\mbox{PEP-II} \xspace}

\def\babar{\mbox{\slshape B\kern-0.1em{\smaller A}\kern-0.1em
  B\kern-0.1em{\smaller A\kern-0.2em R}}\xspace}

\def\invfb  {\ensuremath{\mbox{\,fb}^{-1}}\xspace}

\def\qbar {\ensuremath{\overline q}\xspace}

\def\cbar {\ensuremath{\overline c}\xspace}

\def\b   {\ensuremath{b}\xspace}
\def\bbar {\ensuremath{\overline b}\xspace}

\def \Bd{\mathcal{B}(B_s \rightarrow D_s  X)}
\def \Bl{\mathcal{B}(B_s \rightarrow \ell \nu  X)}
\def \Dp {\mathcal{B}(D_s \rightarrow \phi X)}
\def \Dlp {\mathcal{B}(D_s \rightarrow \ell \nu \phi)}
\def \Dl {\mathcal{B}(D_s \rightarrow \ell\nu X)}

\def\WA{\sum_{i \in u,d,s}{\mathcal{B}(B_s \rightarrow D^{(*)-}_s D_i (X))\mathcal{B}(D_i \rightarrow \ell \nu X)}}

\def\Y#1S{\ensuremath{\Upsilon{(#1S)}}\xspace}

\def\Bbar  {\kern 0.18em\overline{\kern -0.18em B}{}\xspace}

\def\BB   {\ensuremath{B\Bbar}\xspace} 

\def\Kbar {\kern 0.2em\overline{\kern -0.2em K}{}\xspace}

\def\Kz  {\ensuremath{K^0}\xspace}
\def\Kzb  {\ensuremath{\Kbar^0}\xspace}
\def\KzKzb {\ensuremath{\Kz \kern -0.16em \Kzb}\xspace}

\def\result {\ensuremath{9.5^{+2.5 +1.1}_{-2.0 -1.9}}}
\def\resulty {\ensuremath{9.5^{+2.5}_{-2.0}\stat^{+1.1}_{-1.9}\syst}}

\newcommand{\gev}{\ensuremath{{\mathrm{\,Ge\kern -0.1em V}}}\xspace}
\newcommand{\mev}{\ensuremath{{\mathrm{\,Me\kern -0.1em V}}}\xspace}

\newcommand{\gevc}{\ensuremath{{\mathrm{\,Ge\kern -0.1em V\!/}c}}\xspace}
\newcommand{\mevc}{\ensuremath{{\mathrm{\,Me\kern -0.1em V\!/}c}}\xspace}
\newcommand{\gevcc}{\ensuremath{{\mathrm{\,Ge\kern -0.1em V\!/}c^2}}\xspace}
\newcommand{\mevcc}{\ensuremath{{\mathrm{\,Me\kern -0.1em V\!/}c^2}}\xspace}

\newcommand{\stat}{\ensuremath{\mathrm{(stat)}}\xspace}
\newcommand{\syst}{\ensuremath{\mathrm{(syst)}}\xspace}

\allowdisplaybreaks[1]

\begin{document}

\title{
\begin{flushleft}
              \mbox{\textmd{ \normalsize \babar-PUB-11/021 }} 
              \\ \mbox{\textmd{ \normalsize SLAC-PUB-14653 }}
              \\ \mbox{\textmd{ \normalsize arXiv:1110.5600 [hep-ex]}}
       \end{flushleft}
      \vskip 4pt
{
\large \boldmath
A Measurement of the Semileptonic Branching Fraction of the $B_s$ Meson}
}

%
\author{J.~P.~Lees}
\author{V.~Poireau}
\author{V.~Tisserand}
\affiliation{Laboratoire d'Annecy-le-Vieux de Physique des Particules (LAPP), Universit\'e de Savoie, CNRS/IN2P3,  F-74941 Annecy-Le-Vieux, France}
\author{J.~Garra~Tico}
\author{E.~Grauges}
\affiliation{Universitat de Barcelona, Facultat de Fisica, Departament ECM, E-08028 Barcelona, Spain }
\author{M.~Martinelli$^{ab}$}
\author{D.~A.~Milanes$^{a}$}
\author{A.~Palano$^{ab}$ }
\author{M.~Pappagallo$^{ab}$ }
\affiliation{INFN Sezione di Bari$^{a}$; Dipartimento di Fisica, Universit\`a di Bari$^{b}$, I-70126 Bari, Italy }
\author{G.~Eigen}
\author{B.~Stugu}
\affiliation{University of Bergen, Institute of Physics, N-5007 Bergen, Norway }
% ...
\collaboration{The \babar\ Collaboration}
\noaffiliation

\begin{abstract} 

We report a measurement of the inclusive semileptonic branching fraction of the $B_s$ meson using data 
collected with the \babar detector in 
 the center-of-mass (CM) energy region above the $\Upsilon(4S)$ resonance. We use the inclusive yield of $\phi$ mesons and the $\phi$ yield in association
 with a high-momentum lepton to
 perform a simultaneous measurement of the 
 semileptonic branching fraction and the production rate of $B_s$ mesons relative to all $B$ mesons as a function of CM energy. The inclusive
 semileptonic branching fraction of the $B_s$ meson is determined to be $\Bl=\resulty\%$, where $\ell$ indicates the average of $e$ and $\mu$.
 
 \end{abstract}
 
 
\maketitle  

Semileptonic decays of heavy-flavored hadrons serve as a powerful probe of the electroweak and strong interactions and 
are essential to determinations of Cabibbo-Kobayashi-Maskawa (CKM) matrix elements (see, for example, 
``Determination of $V_{cb}$ and $V_{ub}$" in Ref.~\cite{PDG}). The inclusive semileptonic branching fractions of the $B_d$ and $B_u$ mesons
are measured to high precision by experiments operating at the $\Upsilon(4S)$ resonance, which decays almost 
exclusively to \BB\ pairs (here and throughout this note, \BB\ refers to  $B_d \Bbar_d$ and $B_u\Bbar_u$). However, 
lacking an analogous production mechanism,  
information on branching fractions of the $B_s$ meson remains scarce nearly two decades after its first observation~\cite{PDG}. 
Here we report a measurement of the inclusive semileptonic branching fraction of the $B_s$ meson using data collected with 
the \babar detector at the {\pep2}asymmetric-energy electron-positron collider, located at the SLAC National Accelerator Laboratory. The data were collected in a scan of center-of-mass (CM) energies above the $\Upsilon(4S)$ resonance, including the region near the $B_s \Bbar_s$ 
threshold. As $\phi$ mesons are particularly abundant in $B_s$ decays due to the CKM-favored $B_s \to D_s$ transition, the inclusive production rate of $\phi$ mesons and the rate of $\phi$ mesons produced in association with a high momentum electron or muon can be used to simultaneously determine the $B_s$ semileptonic branching fraction and the 
$B_s$ production fraction as a function of the CM energy $E_{\rm CM}$. 

The energy scan data correspond to an integrated luminosity of $4.25\, \invfb$ collected in 2008 in 5\mev steps in the range $10.54\gev 
\le E_{\rm CM} \le 11.2\gev$. In a previous study \cite{Rb}, we presented a measurement of the inclusive $\b$ quark production cross section 
$R_b=\sigma(e^+e^- \rightarrow b\bbar)/\sigma^0(e^+e^- \rightarrow \mu^+\mu^-)$ in this energy range, using this same data sample ($\sigma^0$ is the 
zeroth-order QED cross-section). In the present study, we also make use of $18.55\,\invfb$ of data collected in 2007 at the peak of the \Y4S resonance, and $7.89\,\invfb$ collected 40 \mev below the $\Y4S$, to evaluate backgrounds from continuum ($e^+e^- \to q\qbar, \, q=u,d,s,c$ quark production) and \BB events. We choose below-resonance data for which detector 
conditions most closely resemble those of the scan, and on-resonance data corresponding to roughly twice the 
luminosity of the below-resonance sample. The sizes of these samples are sufficient to reduce the corresponding systematic uncertainties below those associated with irreducible sources.

The \babar detector is described in detail elsewhere~\cite{Aubert:2001tu}. 
The tracking system is composed of a five-layer
silicon vertex tracker (SVT) and a 40-layer drift chamber (DCH) in a
1.5-Tesla axial magnetic field. The SVT provides a precise determination of
the track parameters near the interaction point and standalone tracking for charged particle transverse momenta ($p_t$) down to 50\mevc. The DCH provides a 98\% efficient
measurement of charged particles with $p_t > 500\mevc$. The $p_t$ resolution is $\sigma_{p_t}/p_{t}=(0.13 \cdot p_t +
0.45)\%$.  Hadron and muon identification in \babar\ is achieved by using a likelihood-based
algorithm exploiting specific ionization measured in the SVT and the
DCH in combination with information from an instrumented magnetic-flux return and the Cherenkov angle obtained from the detector
of internally reflected Cherenkov light. Electron identification is provided by a combination of tracking and
information from the CsI(Tl) electromagnetic calorimeter, which also serves to measure photon energies. For the evaluation of
event reconstruction efficiencies across the scan range, simulated samples of $e^+e^- \rightarrow \mu^+\mu^-$, continuum, and $e^+e^- \rightarrow B^{(*)}_{q}\Bbar^{(*)}_q, \, q=u,d,s$ events, created with the \textsc{kk2f} \cite{kk2f}, \textsc{JETSET} \cite{JETSET}, and 
\textsc{EvtGen} \cite{EvtGen} event generators, respectively, are processed through a \textsc{Geant4} \cite{geant4} simulation of the \babar\ detector.

For this measurement, we present the scan data as a function of $E_{\rm CM}$ in bins of 15 \mev. In each bin we measure the number of \BB-like events (defined below), the number of such 
events containing a $\phi$ meson, and the number of events in which the $\phi$ meson is accompanied by a charged lepton candidate. 
The results are normalized to the number of 
$e^+e^- \rightarrow \mu^+\mu^-$ events in the same energy bin so that the luminosity dependence in each bin is removed. These three measurements are
used to extract the fractional number of $B_s\Bbar_s$ events and the semileptonic branching fraction $\Bl$. The procedure is described in detail below.

To suppress QED background, events are preselected with a multihadronic event filter optimized to select \BB and $B_s \Bbar_s$ events. The filter
requires a minimum number of charged tracks in the event ($3$), a minimum total event energy  
(4.5\gev), a well-identified primary vertex near the expected collision point, and a maximum value of the ratio of the second 
to zeroth Fox Wolfram moments \cite{R2} ($R_2 < 0.2$) calculated in the CM frame using both charged tracks
and energy depositions in the calorimeter, where the latter are required not to be associated with a track. 

A different preselection is used to identify muon pair events. 
Events passing this selection must have at least two tracks. The two highest 
momentum tracks are required to be back-to-back in the CM frame to within 10 degrees, appear at large angles to the beam axis ($| \cos{\theta_{\rm CM}} | < 0.7486$), and have an invariant mass greater than 7.5\gevcc. 
In addition, we require that less than 1\gev be deposited in the electromagnetic calorimeter. This selection is 43\% efficient for simulated $\mu^+\mu^-$ events while rejecting virtually all continuum events.

Candidate $\phi$ mesons are reconstructed in the $\phi\to K^+K^-$ decay mode,
by forming pairs of oppositely charged tracks that are consistent with the kaon hypothesis. In each event, the $\phi$ candidate with the best-identified $K^{\pm}$ daughters is selected by assigning 
a weight to each $K^{\pm}$ based on the particle identification criteria. The $\phi$ candidate with the largest sum of kaon weights is selected. The invariant mass distribution of these candidates is used to determine the $\phi$ 
yield in a given $E_{\rm CM}$ bin using a maximum likelihood fit. 
Events containing $\phi$ candidates and an electron or muon 
candidate with a CM momentum exceeding 900\mevc are used to determine the yield of events with both a $\phi$ and a lepton ($\phi$-lepton events). The requirement on the lepton momentum suppresses background from semileptonic charm decays.

\begin{figure}[t]
	\subfigure{\begin{overpic}[trim=0pt 0pt 0pt 0.9cm, clip, scale=.44]{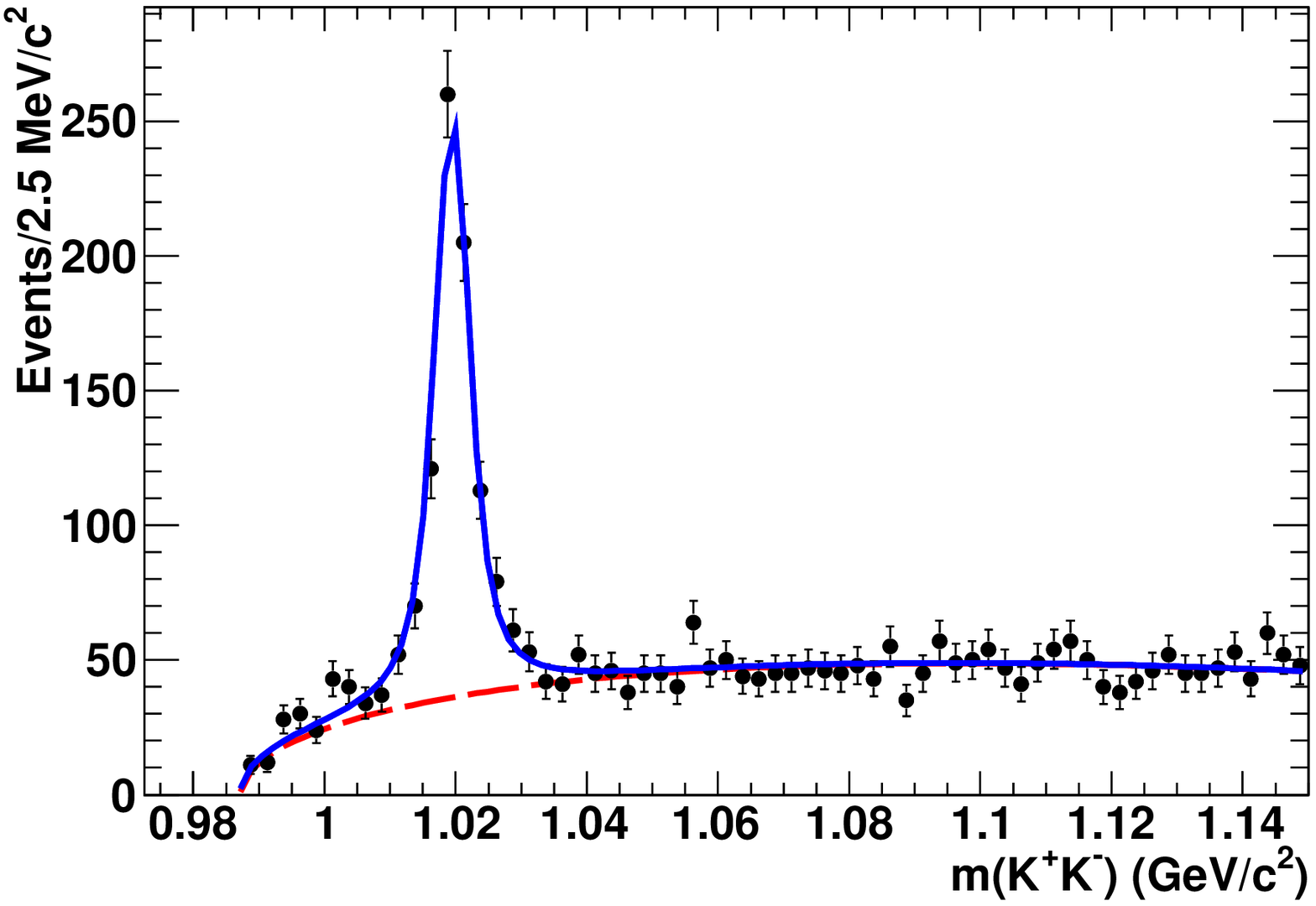}\put(50,55){(a)}\end{overpic}}
	\subfigure{\begin{overpic}[trim=0pt 0pt 0pt 0.9cm, clip, scale=.44]{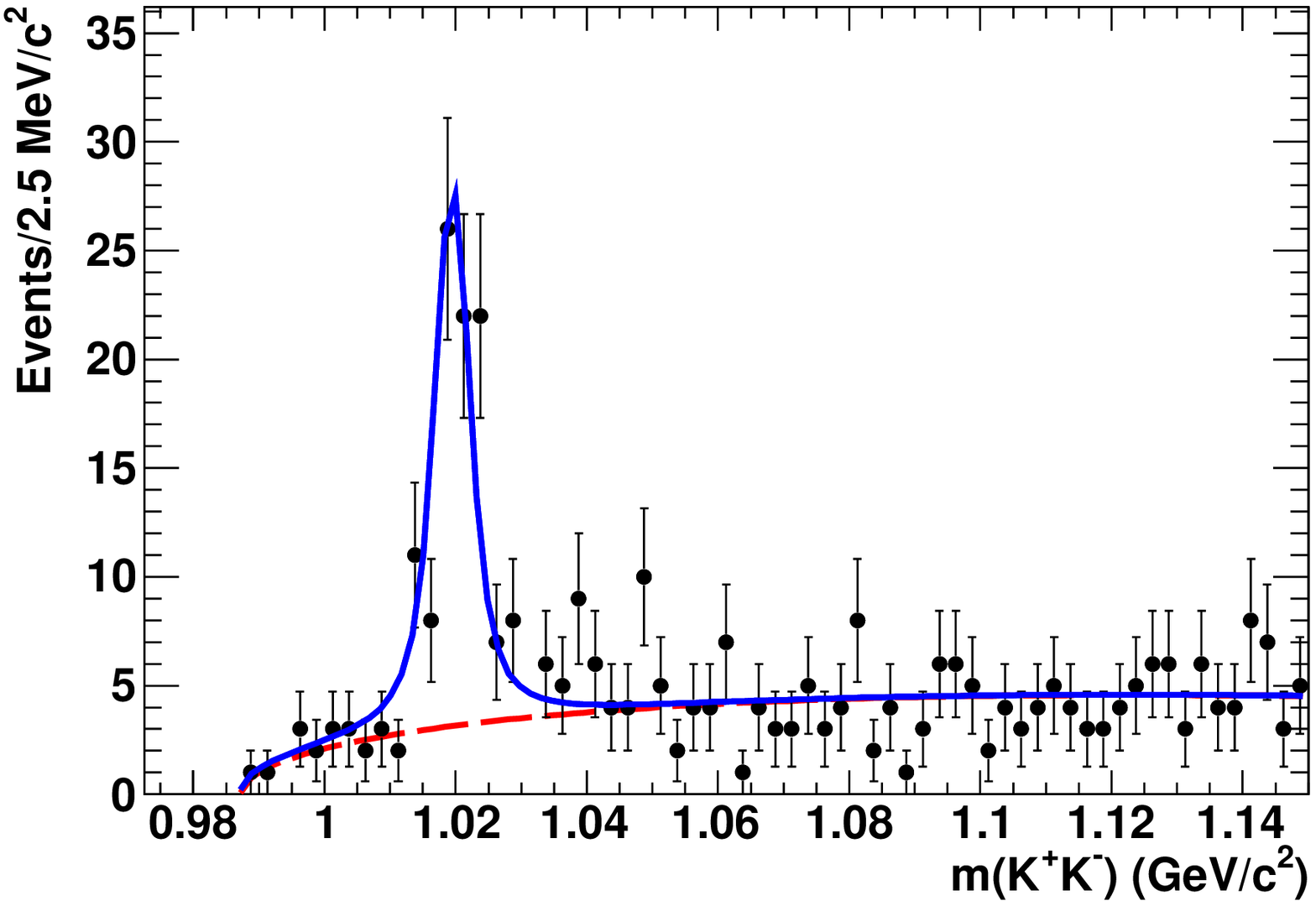}\put(50,55){(b)}\end{overpic}}
\caption[s]{ Invariant mass distribution of $\phi \rightarrow K^+K^-$ candidates in the energy bin 10.8275\gev $\leq E_{\rm CM} \leq$ 10.8425\gev: (a) inclusive $\phi$ candidates; (b) $\phi$-lepton candidates. The background
shape is shown by the dashed curve and the total fit by the solid curve. }
\label{fig:phifits}
\end{figure}

Figure \ref{fig:phifits}
shows, as an example, the $K^+K^-$ invariant mass distribution for (a) all $\phi$ candidates, and (b) $\phi$-lepton candidates, in the energy bin $10.8275 < E_{\rm CM} < 10.8425\gev$. These mass distributions are fit to the function
\begin{equation}
\begin{split}
f(M;N,b,c)\equiv{ }& N V(m_{KK};m_\phi,\Gamma_\phi,\sigma) \\&+ Nc\,(1+b\, m_{KK})\sqrt{1-\left(\frac{2m_K}{m_{KK}}\right)^2},
\label{eqn:fitfunc}
\end{split}
\end{equation}
with $m_K$ the world-average mass value \cite{PDG} of the $K^\pm$. $V(m_{KK};m_\phi,\Gamma_\phi,\sigma)$ is a Voigt
profile (the convolution of a Breit-Wigner function $1/(({m_{KK}}-m_\phi)^2+{\Gamma_\phi}^2/4)$ with a Gaussian resolution function) normalized to unity, so that $N$ 
is the number of events in the peak. We fix the mean ($m_\phi$) and Breit-Wigner width ($\Gamma_\phi$) to the world average values 
of the $\phi$ mass and natural width
 \cite{PDG}, and the width of the Gaussian resolution ($\sigma$) by first performing all of the $\phi$ fits with the parameter left free, then fixing
 it to the weighted mean of all of the values obtained across the scan. The value in data determined by this method is $\sigma = 1.61\pm0.04\stat \mevcc$. 
 The combinatoric background is modeled as the product of a linear term and 
 a threshold cutoff function parameterized by the slope of the linear term ($b$) and a relative scaling ($c$). 

To determine the $\phi$ and $\phi$-lepton yields from $B$ decays in each $E_{\rm CM}$ bin, the contribution of continuum 
events is subtracted. This is achieved by using the data collected below the \Y4S described above.
The event, $\phi$, and $\phi$-lepton yields are measured in this dataset following the same procedures described above. These yields are corrected for the 
 energy dependence of the reconstruction efficiencies and are then subtracted from the scan yields in each $E_{\rm CM}$ bin. 
 This procedure neglects the 
 different energy dependence of a small component of the hadronic and dimuon cross sections, primarily due to the presence of initial state radiative (ISR) 
 $e^+e^-\rightarrow\gamma\Upsilon(1S,2S,3S)$ and two photon 
 $e^+e^- \rightarrow e^+e^-\gamma^{*}\gamma^{*} \rightarrow e^+e^- X_h$ events, which do not scale according to 
 $1/E^2_{\rm CM}$. The effect of these contributions is to introduce a small energy dependence on the amount to be subtracted from each bin. The average size of this 
 effect is estimated to be less than 2\% of the below-resonance event yield. The impact on the result is taken as a systematic uncertainty. 

\begin{figure}[bht]
\centering
\subfigure{\begin{overpic}[trim=0pt 0pt 0pt 0.55cm, clip,scale=0.46]{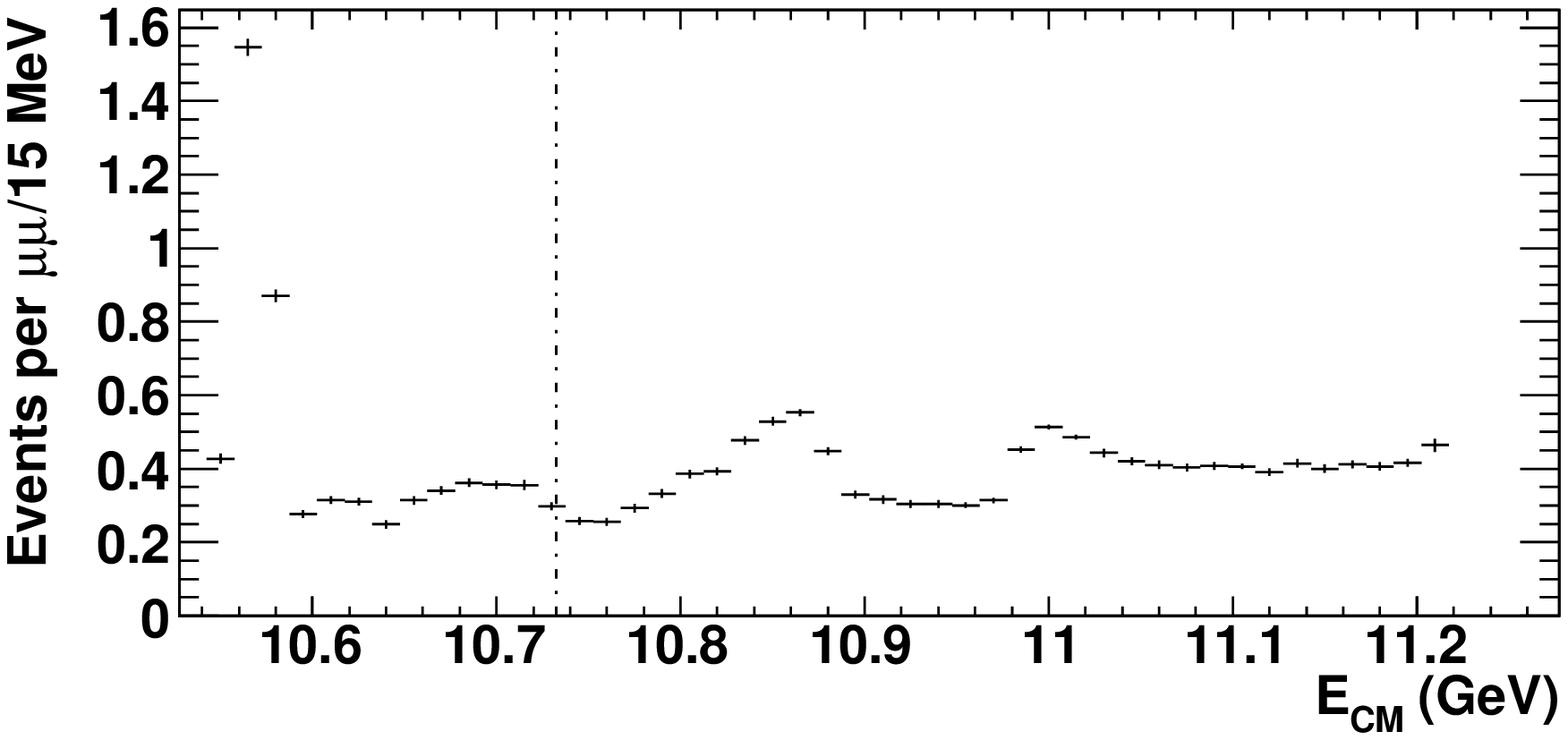}\put(50,38){(a)}\end{overpic}}
\subfigure{\begin{overpic}[trim=0pt 0pt 0pt 0.55cm, clip,scale=0.46]{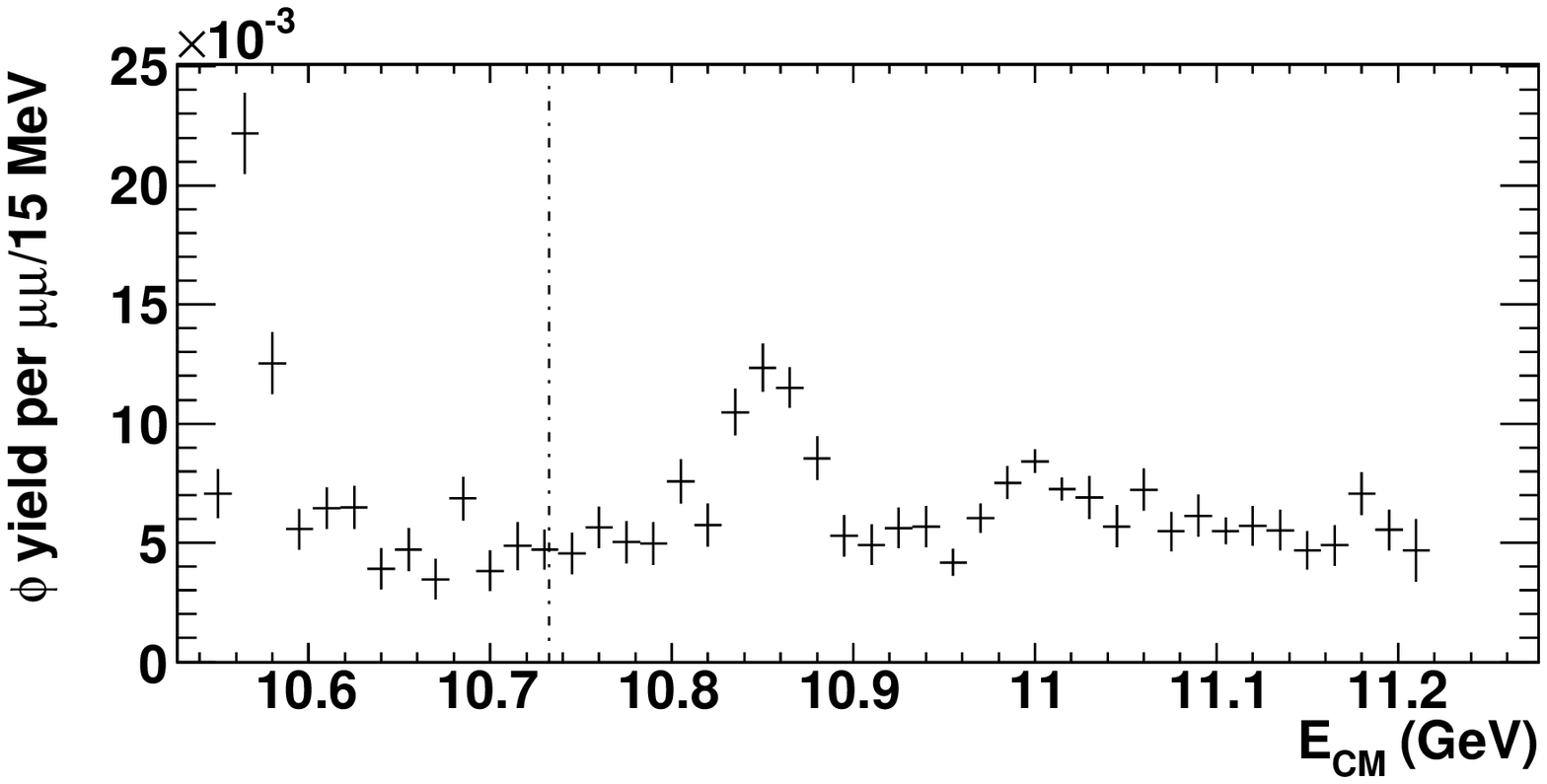}\put(50,38){(b)}\end{overpic}}
\subfigure{\begin{overpic}[trim=0pt 0pt 0pt 0.55cm, clip,scale=0.46]{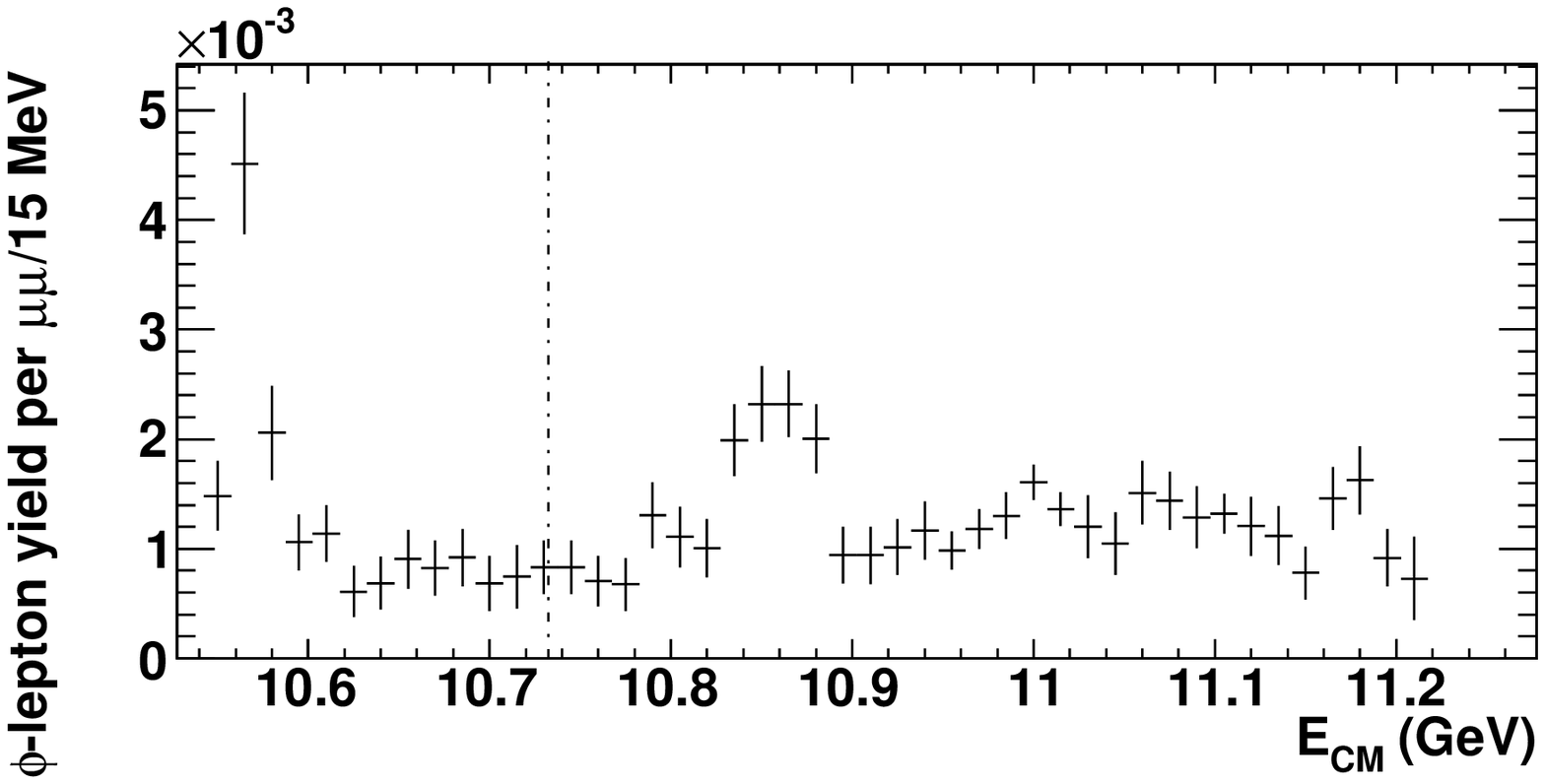}\put(50,38){(c)}\end{overpic}}
\caption{Relative (a) event,  (b) $\phi$, and (c) $\phi$-lepton yields, normalized to the $\mu^+\mu^-$ yields. Corrections for detector efficiency have not been applied. The dotted vertical line indicates the $B_s$ production threshold.}
\label{fig:counts}
\end{figure}

The normalized event, $\phi$, and $\phi$-lepton yields after the continuum 
subtraction are presented in Fig. \ref{fig:counts}. These three quantities, denoted $C_h$, $C_{\phi}$ and 
$C_{\phi\ell}$ respectively, can be expressed in terms of contributions from events containing $B^{(*)}_{u,d}$ and $B^{(*)}_s$ events, the cross section ratio $R_B \equiv  \sum\limits_{q=\{u,d,s\}}{\sigma(e^+e^- \to B_q\Bbar_q})/{\sigma_{\mu^+\mu^-}}$, and the related reconstruction efficiencies, as follows:
\begin{flalign}
C_h = R_B&\left[ f_s \epsilon^s_h + (1-f_s) \epsilon_h \right]\label{eqn:c1}\displaybreak[0]\\
C_{\phi} = R_B& \left[ f_s \epsilon^s_\phi P(B_s \Bbar_s \rightarrow \phi X) \right. \nonumber\\
		& \left. +(1-f_s) \epsilon_\phi P(B \Bbar \rightarrow \phi X)\right]\label{eqn:c2}\displaybreak[0]\\
C_{\phi\ell} \!\! = R_B& \left[ f_s \epsilon^s_{\phi\ell} P(B_s \Bbar_s \rightarrow \phi \ell X) \right. \nonumber \\
		& \left. +(1-f_s) \epsilon_{\phi\ell} P(B \Bbar \rightarrow \phi \ell X)\right] \label{eqn:c3}
\end{flalign}
(with energy dependence implicit in all terms here and elsewhere), where 
\begin{equation}
f_s \equiv \frac{N_{B_s}}{N_{B_u} + N_{B_d}+N_{B_s}}
\end{equation}
and $\epsilon_X$($\epsilon^s_X$) is the efficiency for a $B_{u,d}$ ($B_s$) pair to 
contribute to the event, $\phi$ or $\phi$-lepton yield. The efficiencies are estimated from simulation, while 
$P(B\Bbar \rightarrow \phi{ \rm X})$ and $P(B\Bbar \rightarrow \phi \ell { \rm X})$, 
which are the probabilities that a $\phi$ or a $\phi$-lepton combination is produced in an event with a $\BB$ pair,
are measured using the $\Y4S$ data sample described above. Specifically, we determine the  $\phi$ and $\phi$-lepton yields in the the $\Y4S$ data. We then apply Eqs. \eqref{eqn:c1}, \eqref{eqn:c2}, and \eqref{eqn:c3} with $f_s = 0$ to extract $\epsilon_{\phi} P(B\Bbar \rightarrow \phi{ \rm X})$ and $\epsilon_{\phi\ell} P(B\Bbar \rightarrow \phi \ell { \rm X})$. Simulations are used to extrapolate the values of the efficiencies to other energies. 

\begin{figure}[htb]
\includegraphics[trim=0pt 0pt 0pt 1.0cm, clip,scale=.45]{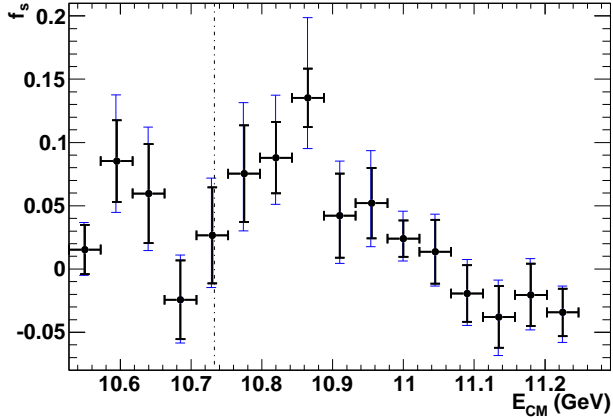}
\caption{Results for the fraction $f_s$ as a function of $E_{\rm CM}$. The inner error bars show the statistical uncertainties and the outer error bars the statistical and systematic uncertainties added in quadrature. The dotted line denotes the $B_s$ threshold.}
\label{fig:fs}
\end{figure}

The remaining unknown quantities of interest are the probabilities $P(B_s\Bbar_s \rightarrow \phi { \rm X})$ and $P(B_s\Bbar_s \rightarrow \phi \ell { \rm X})$ that a $B_s \Bbar_s$ pair will yield a $\phi$ or $\phi$-lepton event. To estimate $P(B_s\Bbar_s \rightarrow \phi X)$ we use the 
current world averages \cite{PDG} of the inclusive branching fractions 
$\Bd$, $\Dp$, and $\mathcal{B}(D \rightarrow \phi X)$. Here and in the following $D$ refers to the sum of $D^{\pm}$ and $D^0$ contributions. Also needed are estimates of the unmeasured branching fractions 
$\mathcal{B}(B_s \rightarrow c\cbar\phi)$ and $\mathcal{B}(B_s \rightarrow D D_s X)$. The former quantity accounts for direct $B_s \to \phi$ production, a substantial 
fraction of which arises from $B_s$ to charmonium decays. We use the central value from the simulation, $1.7\%$, which is roughly consistent with charmonium 
production in the $B$ system. For the 
latter quantity we use a naive quark model prediction of 15\% for $b\to ccs$. 

The inclusive $\phi$ yield in $B_s$ decays can be expressed as:
\begin{equation}
\begin{split}
P(B_s\rightarrow \phi X) &= \mathcal{B}(B_s \rightarrow D^{(*)}_s X) \;
 \mathcal{B}(D_s \rightarrow \phi X) \\
&+ \mathcal{B}(B_s  \rightarrow  c\cbar\phi)\\
&+\mathcal{B}(B_s \rightarrow D D_s X)\;
\mathcal{B}(D\rightarrow \phi X),
\end{split}
\label{eqn:PBsBs}
\end{equation}
from which we determine
\begin{equation}
P(B_s\Bbar_s\rightarrow \phi X)=2P(B_s\rightarrow \phi X) - P(B_s\rightarrow \phi X)^2.
\label{eqn:PBsBs1}
\end{equation}
The unknown quantities in Eqs. \eqref{eqn:c1} and \eqref{eqn:c2} are $f_s$ and the common normalization $R_B$. The ratio $f_s$ can be determined as a function of $E_{\rm CM}$ by eliminating $R_B$ between the two equations. The result is presented in Fig. \ref{fig:fs}.
The ratio $f_s$ peaks around the $\Upsilon(5S)$ mass. The total excess below the $B_s \Bbar_s$ threshold and deficit above 11 \gev are consistent with zero within 1.5 and 1.3 standard deviations, respectively. 

Using Eq.~\eqref{eqn:c3}, a $\chi^2$ is constructed from the measured and expected values of $P(B_s\Bbar_s \rightarrow \phi \ell X)$
across the entire scan. 
The $\chi^2$ is minimized with respect to $\Bl$. The following processes contribute to $C_{\phi\ell}$ from $B_s\Bbar_s$ events: 
primary leptons originating from a $B_s$ semileptonic decay,
secondary leptons resulting from semileptonic decays of charmed mesons,
and $\pi^\pm$ or $K^\pm$ misidentified as $e^\pm$ or $\mu^\pm$.
The contribution from primary leptons arises from events where one or both $B_s$ mesons decay semileptonically, 
and we determine the $\phi$-lepton efficiency for each case (denoted $\epsilon^s_{\phi\ell}$ for one semileptonic decay and $\epsilon^s_{\phi\ell\ell}$ for two). It is found that $\epsilon^s_{\phi\ell}$ ranges from $8.5\%-10\%$ and $\epsilon^s_{\phi\ell\ell}$ is about $10\%$. 

 \begin{figure}[tbh]
{\includegraphics[trim=0pt 0pt 0pt 1cm, clip,scale=.44]{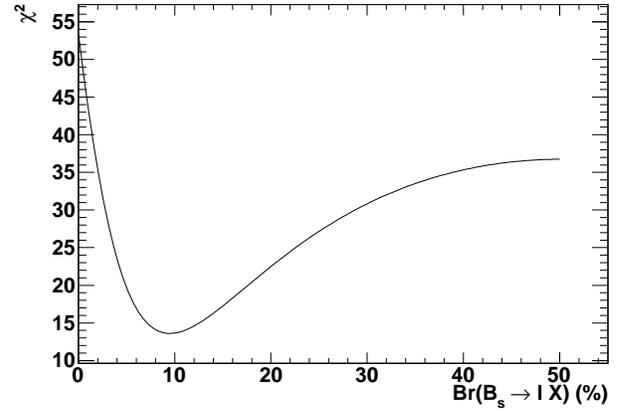}}
\caption{$\chi^2$ formed from the measured and expected yields, as described in the text, as a function of the semileptonic branching fraction. Note that since we express the branching fraction as the average of the $e$ and $\mu$ channels, the physical bound is 50\%.}
\label{fig:chi2}
\end{figure}

For the secondary lepton contribution, we consider events with up to two leptons 
coming from $D^\pm, D^0$ or $D^\pm_s$ decays. The selection efficiency in this case is estimated as the product of the $\phi$ reconstruction efficiency in $B_s \Bbar_s$ events in which neither $B_s$ decays semileptonically but a lepton candidate is identified (referred to below as $\epsilon^D_\phi$), and
a lepton detection efficiency determined from simulation ($\epsilon^D_\ell$). It is found that $\epsilon^D_\phi$ lies in the range $15\%-16.5\%$, and $\epsilon^D_\ell$ in the range $8\%-9.5\%$ per lepton.
The contribution 
from hadrons that are misidentified as leptons is estimated from simulation to be 3.3\% of the $\phi$-lepton 
candidates in $B_s\Bbar_s$ events. 

For the expected and measured $\phi$ yields, we find:
\onecolumngrid
\begin{align}
\begin{split}
\epsilon^s_{\phi\ell}P(B_s\Bbar_s \rightarrow \phi \ell X)_{\rm Primary}={ }&(2\epsilon^s_{\phi\ell} - \epsilon^s_{\phi\ell\ell}) \mathcal{B}(D_s \rightarrow \phi  X) \left[ -2 + \mathcal{B}(D_s \rightarrow \phi  X) \right][\Bl]^2\\
&+\Bl \epsilon^s_{\phi\ell} \Big[\mathcal{B}(D_s \rightarrow \phi  X) + [1-\mathcal{B}(D_s \rightarrow \phi  X)] P(B_s \rightarrow \phi X)\Big],\label{eqn:primary}\qquad
\end{split}\\
\begin{split}
\epsilon^s_{\phi\ell}P(B_s\Bbar_s \rightarrow \phi \ell X)_{\rm Secondary}={ }&2 \epsilon^D_\ell \epsilon^D_\phi \Bigg\{  \Big[ \Dlp + \Dl \Dp \Big. \Bigg.  \\
&- \Big.\Dlp \Dp \Big]   [\Bl]^2 \\
&+\Big[P(B_s \rightarrow \phi X)(\Dlp - \Dl) - \Dlp \Big. \\
&- \Bd \Dlp - \Bd \Dl \Dp \\
&+ \Bd \Dlp \Dp \\
&+ \Big. (\Dp - 2) \WA \Big] \Bl \\
&+ \Bd P(B_s \rightarrow \phi X) \left[ \Dl - \Dlp \right]\\
&+ \Bd \Dlp \\
&+ \left[  \mathcal{B}(B_s \rightarrow \phi X) + \Dp -  P(B_s \rightarrow \phi X) \Dp \right] \\
&\times \Bigg. \WA \Bigg\},\label{eqn:secondary}
\end{split}\\
\begin{split}
\epsilon^s_{\phi\ell}P(B_s\Bbar_s \rightarrow \phi \ell X)_{\rm Expected} ={ }&\left\{ \epsilon^s_{\phi\ell} \times 0.591 \times \Bl - (2\epsilon^s_{\phi\ell}-\epsilon^s_{\phi\ell\ell}) \times 0.289 \times (\Bl)^2 \right.\\
&\left.+\epsilon^D_\phi \epsilon^{D}_\ell \left[ 0.1375 - 0.2721 \times \Bl + 0.1339 \times (\Bl)^2 \right]\right\}, \label{eqn:expected}
\end{split}\\
\begin{split}
\epsilon^s_{\phi\ell}P(B_s\Bbar_s \rightarrow \phi \ell X)_{\rm Measured} ={ }&(1-0.033)\left( C_{\phi\ell}\frac{f_s\epsilon^s_h + (1-f_s)\epsilon_h}{f_s C_h} - \frac{(1-f_s)\epsilon_{\phi\ell}P(B\Bbar \rightarrow \phi \ell X)}{f_s}\right),\qquad\quad\label{eqn:measured}
\end{split}
\end{align}
\twocolumngrid
\noindent where that Eq.~\eqref{eqn:expected} is the sum of Eqs.~\eqref{eqn:primary} and \eqref{eqn:secondary} after substituting the values of known quantities. The first line in Eq.~\eqref{eqn:expected} expresses the contribution from primary leptons and the second that from secondary leptons. Eqs. \eqref{eqn:expected} and \eqref{eqn:measured} describe the measured and expected values used to form the $\chi^2$ to be minimized, along with the statistical uncertainties of each of the measured quantities and the uncertainties in the energy-dependent efficiencies.

The expression in Eq.~\eqref{eqn:expected} for the expected value of $\epsilon^s_{\phi\ell}P(B_s\Bbar_s \rightarrow \phi \ell X)$ is quadratic in the unknown $\Bl$, and so a $\chi^2$ formed from the deviation of the expected from the measured values, summed over all bins above the $B_s\Bbar_s$ threshold, is quartic in this unknown. 
Minimizing the $\chi^2$ with respect to the $B_s$ semileptonic branching fraction we find $\Bl = 9.5^{+2.5}_{-2.0}\%$. Figure \ref{fig:chi2} shows the dependence of $\chi^2$ on $\Bl$.

Systematic uncertainties are summarized in Table
\ref{tab:syst} and include the contributions described below. 

\begin{itemize}
\addtolength{\itemsep}{-0.6\baselineskip}

\item Uncertainties for branching fractions, which are either taken from Ref. \cite{PDG}  when known, or assumed to be 50\% for $\mathcal{B}(B_s \rightarrow c\cbar\phi)$ and $\mathcal{B}(B_s \rightarrow D D_s X)$. These are separately listed in Table \ref{tab:syst}, as is $\Bd$, which contributes a very large uncertainty compared to the other branching fractions.
\item Requirements used in the event preselection, including the lepton momentum requirement. The uncertainty due to the lepton momentum requirement dominates in this group, and reflects the dependence of the result on the decay model used to simulate $B_s$ semileptonic decays.
\item Fixed parameters used in the fits to $m_{KK}$, including $m_\phi$, $\Gamma_\phi$, $\sigma$. 
\item The parameterization of the background and absence of a term in the fit corresponding to threshold contributions from light scalars.
\item Uncertainties in particle identification (PID) efficiencies and hadron misidentification probabilities.
\item The determination of $P(B\Bbar\rightarrow \phi X)$ and $P(B\Bbar\rightarrow \phi \ell X)$ in $\Y4S$ data (these quantities are determined to 1\% and 1.8\% relative uncertainty).
\item Sensitivity of efficiencies to differences in branching fractions implemented in simulation compared to their measured values.
\item Uncertainties in the continuum-subtracted number of events due to ISR and two photon events, which do not follow a 1/$E_{\rm CM}^2$ energy dependence.
\item A correction made to the continuum subtraction of the number of $B\Bbar$-like events due to an over-subtraction found in simulation studies. The size of this correction is about 1\% of the amount to be subtracted; we use $\pm100\%$ of this correction as a systematic uncertainty. 
\item Possible bias in the $\chi^2$ minimization technique  at low statistics. Firstly, evaluating the behavior of this method for extracting $\Bl$ for many pseudo-data samples derived from the simulated dataset gives evidence for a small bias at low statistics. Secondly, it was found that the analysis performed in high statistics simulation tends to overestimate $\Bl$ by an amount corresponding to half the statistical error reported.
 \end{itemize}
 
To determine whether the uncertainties from these sources scale with the result or not, each was evaluated in a simulation sample with a higher semileptonic branching fraction and compared with the result in the normal simulation sample.
It was found that the uncertainty from the determination of $P(B\Bbar\rightarrow \phi \ell X)$ in $\Y4S$ data does not scale with the branching fraction, nor does the uncertainty contributed by several of the input branching fractions. These are thus separated in Table \ref{tab:syst}. The remaining uncertainties are found to scale with $\Bl$ and thus to be multiplicative.

\begin{table}[t]
\caption{Relative multiplicative and additive systematic uncertainties for the measurement of $\Bl$.}
\begin{ruledtabular}
\begin{tabular}{ p{0.60\columnwidth}  p{0.40\columnwidth} }
{\bf Multiplicative Systematics} &{\bf Relative \par Uncertainty (\%)}\\\hline
$\mathcal{B}(B_s \to D^{(*)}_s X)$ & ${+8.72}/{-13.58}$\\
$\mathcal{B}(B_s \to c\cbar\phi)$ (Unmeasured) & ${\pm 3.20}$\\
$\mathcal{B}(B_s \rightarrow D D_s X)$ (Unmeasured) & ${+1.12}/{-1.16}$\\
Other Branching Fractions & ${+0.52}/{-0.54}$\\
Event and Lepton Selection & ${+1.99}/{-2.85}$\\
Fixed Fit Parameters & ${+0.49}/{-0.15}$\\
Background Parameterization & ${\pm 0.93}$\\
PID and Lepton Fake Rate & $\pm 3.21$\\
$P(B_{u,d}\Bbar_{u,d} \to \phi)$ & ${+1.47}/{-1.69}$\\
Simulation Branching Fractions & $\pm2.59$\\
ISR and 2$\gamma$ Background & ${+1.57}/{-7.14}$\\
{Correction to Event Subtraction } & ${+1.88}/{-4.59}$\\
Technique bias & ${+0.39}/{-10.00}$\\\hline
{\bf Total Multiplicative} & $({+10.87}/{-19.92})\%$\\\hline
{\bf Additive Systematics} &{\bf Uncertainty ($\times 10^{-3}$)}\\\hline
Other Branching Fractions & ${+0.56}/{-0.64}$\\
$P(B_{u,d}\Bbar_{u,d} \to \phi \ell \nu)$ & ${+4.30}/{-3.90}$\\\hline
{\bf Total Additive} & $({+4.34}/{-3.95})\times 10^{-3}$\\\hline
{\bf Total Systematic} & $({+11.20}/{-19.34})\!\!\times\!\! 10^{-3}$
\end{tabular}
\end{ruledtabular}
\label{tab:syst}
\end{table}

Our final result for the inclusive semileptonic branching fraction is $\result\%$, which is the average of the branching fractions to $e$ and $\mu$.

In conclusion, we performed a simultaneous measurement of the $B_s$ semileptonic branching fraction and its production rates in the CM energy region from 10.56\gev to 11.20\gev. The semileptonic branching fraction is consistent with theoretical calculations in Refs. \cite{Gronau} and \cite{Beauty}. Our measurement of the $B_s$ production rates are consistent with the predictions of coupled channel models \cite{tornqvist}, in which $B_s$ production peaks near the $\Upsilon(5S)$ and is vanishingly small elsewhere.

We are grateful for the excellent luminosity and machine conditions
provided by our \pep2 colleagues, 
and for the substantial dedicated effort from
the computing organizations that support \babar.
The collaborating institutions wish to thank 
SLAC for its support and kind hospitality. 
This work is supported by
DOE
and NSF (USA),
NSERC (Canada),
CEA and
CNRS-IN2P3
(France),
BMBF and DFG
(Germany),
INFN (Italy),
FOM (The Netherlands),
NFR (Norway),
MES (Russia),
MICIIN (Spain),
STFC (United Kingdom). 
Individuals have received support from the
Marie Curie EIF (European Union),
the A.~P.~Sloan Foundation (USA)
and the Binational Science Foundation (USA-Israel).

\end{document}